\documentstyle[fleqn,epsf,11pt]{article}
 
\def\bo{{\raise.15ex\hbox{\large$\Box$}}}

\def\dag{^{\dagger}{}}

\def\ordless{{\lower2mm\hbox{$\,\stackrel{\textstyle <}{\sim}\, $}}}
\def\ordmore{{\lower2mm\hbox{$\,\stackrel{\textstyle >}{\sim}\, $}}}

\newtoks\slashfraction
\slashfraction={.13}
\def\slash#1{\setbox0\hbox{$\, #1$}
\setbox0\hbox to \the\slashfraction\wd0{\hss \box0}/\box0}

\def\leftrightarrowfill{$\mathsurround=0pt \mathord\leftarrow \mkern-6mu
        \cleaders\hbox{$\mkern-2mu \mathord- \mkern-2mu$}\hfill
        \mkern-6mu \mathord\rightarrow$}
\def\overleftrightarrow#1{\vbox{\ialign{##\crcr
        \leftrightarrowfill\crcr\noalign{\kern-1pt\nointerlineskip}
        $\hfil\displaystyle{#1}\hfil$\crcr}}}
\def\startarray{\left( \begin{array}}
\def\finarray{\end{array} \right)}
\def\starteq{
\begin{eqnarray}}
\def\fineq{\end{eqnarray}
}

\catcode`@=11
\def\underline#1{\relax\ifmmode\@@underline#1\else
$\@@underline{\hbox{#1}}$\relax\fi}
\catcode`@=12
 
\def\to{\mbox{-}}

\newskip\humongous \humongous=0pt plus 1000pt minus 1000pt

\newif\ifdtup
 
\def\textcite#1{Ref.~{\cite{#1}}}

\def\thefootnote{\fnsymbol{footnote}}
 
\def\author#1#2{{\bf #1} \\ {\em #2}\vspace{5mm}}

\def\bold#1{\setbox0=\hbox{$#1$}%
     \kern-.025em\copy0\kern-\wd0
     \kern.05em\copy0\kern-\wd0
     \kern-.025em\raise.0433em\box0 }

\def\title#1#2#3#4#5{\thispagestyle{empty}
        \begin{center} \vspace*{1cm} { \bf #3} \\[.5in] {#4{}}
        \end{center} \vfill \centerline{ ABSTRACT}
   {\nopagebreak \noindent\begin{quotation}\noindent {\small #5}
   \end{quotation}} \vfill {#2} \hfill\begin{tabular}{r} {#1} 
        \end{tabular}  \newpage
        \def\thefootnote{\arabic{footnote}}}
\def\prefer{\section*{}
    \list{[\arabic{enumi}]}{\usecounter{enumi}\settowidth\labelwidth{[000]}
      \leftmargin\labelwidth\advance\leftmargin\labelsep \rightmargin=0pt}
        \small \sfcode`\.=1000\relax}

\def\refer#1{\section*{\large \sc {#1}}
    \list{\arabic{enumi}.}{\usecounter{enumi}\settowidth\labelwidth{[000]}
      \leftmargin\labelwidth\advance\leftmargin\labelsep \rightmargin=0pt}
        \raggedright \small \sfcode`\.=1000\relax}

\def\ReFer#1#2{\section*{\large\sc#1}
    \list{[\arabic{enumi}]}{\usecounter{enumi}\settowidth\labelwidth{#2} 
      \leftmargin\labelwidth\advance\leftmargin\labelsep \rightmargin=0pt}
        \raggedright \small \sfcode`\.=1000\relax}

\def\REFER#1#2{\section*{\large\sc#1}
    \list{#2 {enumi}.}{\usecounter{enumi}\settowidth\labelwidth{[000]}
      \leftmargin\labelwidth\advance\leftmargin\labelsep \rightmargin=0pt}
        \raggedright \small \sfcode`\.=1000\relax}

\def\startbib{\vspace{1in}\begin{refer}{References}
\small\frenchspacing\nopagebreak}
\def\endbib{\end{refer} \normalsize \nonfrenchspacing}
\def\startfig{\newpage \centerline{{\sl Figure captions}} \begin{itemize}}

\def\endfig{\end{itemize}}
\newcommand{\be}{\begin{equation}}
\newcommand{\ee}{\end{equation}}
\newcommand{\bea}{\begin{eqnarray}}
\newcommand{\eea}{\end{eqnarray}}

\newcommand{\AmS}{{\protect\the\textfont2
  A\kern-.1667em\lower.5ex\hbox{M}\kern-.125emS}} 
\begin{document}

\title{October 2002} {PC073.1002}
{Charming penguin and direct  CP-violation  in  
charmless $B$ decays   
\footnotetext{${\dag}$  
Talk given at the QCD Euroconference 2002, Montpellier 2-9 July 2002}}
{\author{T. N. Pham} {Centre de Physique Th\'eorique, \\
Centre National de la Recherche Scientifique, UMR 7644, \\  
Ecole Polytechnique, 91128 Palaiseau Cedex, France}} 
{In the study of two-body charmless $B$ decays as a mean of looking for
direct CP-violation and measuring the CKM mixing parameters in the
Standard Model, the short-distance penguin contribution 
with its absorptive part generated by charm quark loop
seems capable of producing sufficient $B \rightarrow K\pi$ decays rates, as
obtained in factorization and QCD-improved factorization models. However
there are also long-distance  charming penguin contributions which could
give rise to a strong phase due to  the rescattering 
$D^{*}D^{*}_{s} \rightarrow K\pi$ etc. In this talk, I would like
to discuss recent works  on the 
charming penguin contribution as 
a different approach to the calculation of these contributions
in two-body charmless $B$ decays. We find that the charming penguin 
contribution is significant for $B\rightarrow K\pi$ decays and, 
together with the tree and penguin terms,  produces
large branching ratios in agreement with data, though the
analysis is affected by large theoretical uncertainties. The absorptive
part due to the charmed meson intermediate states is found to produce
large CP asymmetries for $B \rightarrow K\pi,\,\pi\pi$ decays.}

In nonleptonic charmless two-body $B$ decays, the interference between 
the tree-level and penguin terms
make it possible to look for direct CP-violation and 
to determine the CP-violating weak CKM phase angle $\gamma$ from these
decays \cite{Cleo,Babar,Belle}. It is known
that the top quark penguin amplitude is insufficient 
to produce a large $B \rightarrow K\pi$ decay rates. 
When the effects of charm quark loop are included, the decay rate is 
enhanced \cite{Gerard,Fleischer,Fleischermannel,Deshpande1,Ali} 
and are in qualitative agreement with 
data \cite{Ali,Deandrea,Deshpande,Isola1}. The effects of charm quark 
loop is contained in QCD-improved factorization as shown
recently \cite{Beneke,Du,Muta}.

One can also take the view
that these charm quark loop contributions are basically long-distance
effects,  essentially due to charmed meson rescattering processes, such
as, e.g. $B\rightarrow D_s D\rightarrow K\pi$. 
 These contributions, first
discussed  in \cite{Nardulli}, have been more recently
stressed by \cite{Ciuchini1,Ciuchini2} where they are called the 
charming penguin contributions. In \cite{Isola2,Isola} , we have evaluated
these contributions and found that both the absorptive and the real part
of the charming penguin are large, comparable to the short-distance part
and is capable of generating large CP asymmetries in charmless $B$ decays.
In this talk I would like to discuss these calculations 
and stress the importance of the charming penguin in charmless $B$
decays, especially the large absorptive part generated by the 
charmed meson rescattering effects and the CP asymmetries obtained. 
That the charmed meson rescatterings could generate a large charming
penguin is easily seen from the unitarity of the $S$--matrix.  
For example, the absorptive part of the $B \rightarrow K\pi$ decays amplitude
would get the main contribution from the $D_{s}D$ and $D_{s}^{*}D^{*}$
intermediate states since the color-favored, Cabibbo-allowed
$B \rightarrow D_{s}D$  and $B \rightarrow D_{s}^{*}D^{*}$ decays which are governed
by the large coefficient $a_{2} = 1.03$ have 
branching ratios of the order $10^{-2}$ compared with the penguin-dominated
$B \rightarrow K\pi$ decays with branching ratios a few $10^{-5}$. Without
the suppression of the process $D_{s}D \rightarrow K\pi$ and 
$D_{s}^{*}D^{*} \rightarrow K\pi$,
we would have a very large $B \rightarrow K\pi$ decay rates. 

In the standard model, the effective Hamiltonian for $B \rightarrow K\pi$ decays
are given by 
\bea
 H_{\rm eff} &=& {G_F\over \sqrt{2}}[V_{ub}V^*_{us}(
c_1O_{1}^{u} + c_2O_{2}^{u})  + V_{cb}V^*_{cs}(c_1O_{1}^{c} \nonumber\\
&&  + \  c_2 O_{2}^{c}) 
-V_{tb}^* V_{ts} \sum_{i=3}^{10} c_i O_i + c_g O_g] + {\rm h.c.}\;,
\label{hw}
\eea
where $c_i$ are next-to-leading Wilson coefficients at the normalization 
scale $\mu = m_b$ \cite{Fleischer,Deshpande1,Buras,Ciuchini,Kramer}.
$V_{ub}$ etc. are elements of the unitary Cabibbo-Kobayashi-Maskawa
(CKM) quark mixing matrix relating the  weak interaction 
eigenstate of $d,s,b$ quark to their mass eigenstate \cite{Nir}. The
CP-violating weak phase $\gamma = arg(-V_{ud}V_{ub}^{*}/V_{cd}V_{cb}^{*}) $
produces  CP asymmetries in charmless $B$ decays.
The tree-level operators $O_{1}^{u}, O_{2}^{u} $, the penguin
operators $O_i$ ($i=3,..., 10$) and the chromomagnetic gluon operator
$O_g$ give the short-distance amplitude. The operators
$O_{1}^{c}, O_{2}^{c} $ contribute to the $B \rightarrow K\pi$ decay amplitudes
only through loop generated by the charmed meson 
intermediate states
with the absorptive part given by the process $B \rightarrow D_{s}D \rightarrow K\pi$
and $B \rightarrow D^{*}_{s}D^{*} \rightarrow K\pi$ and the real part by the following
expression  based on the light-cone expansion \cite{Wilson,Brandt} 
\begin{eqnarray}
&& \kern -0.7cm  A_{LD}= K\langle K^0\pi^+|:J_{\mu}(0)\hat
  J^{\mu}(0):|B^+\rangle\cr&&\approx  K
  \int\frac{d\vec n}{4\pi}\langle K^0\pi^+|T(J_{\mu}(x_0)\hat
  J^{\mu}(0))|B^+\rangle
\label{eq:1}
\end{eqnarray}
with $J_{\mu} = \bar{b}\gamma_{\mu}(1 - \gamma _5)c$ and $\hat
J_{\mu} = \bar{c}\gamma_{\nu}(1 - \gamma _5)s$ ;
$a_2 =(c_2+c_1/3)\, (=1.03)$ 
$c_1$ and $c_2$ are Wilson coefficients, and $K= {G_F \over \sqrt{2}} \,
a_2\, V_{cb}^* V_{cs}$. $x_0=(0,\,{\vec n}/{\mu})$ with  
$|\vec n|=1$ and $\mu $  sufficiently large ($\mu\sim m_b$) so that 
$O(x_{0}^{2})$ terms become negligible. $\mu$ acts as a separation between
short-distance physics (in the Wilson coefficients) and the long-distance
physics which is dominated by the hadronic states and resonances. We have,
\be
A_{LD} =K\int\frac{d^4 q}{(2\pi)^4}\, \theta(q^2 + \mu^{2})\,T(q)
\label{eq:3}
\ee
$T(q)~=~\int \,d^{4}x  \, e^{+iq\cdot x} \langle K^0\pi^+|{\rm T}
(J^{\mu}(x)\hat J_{\mu}(0))|B^+\rangle $ and the high frequency part has been
cut-off from the q integral by  $\theta(q^2 + \mu^{2}) $ 
to simplify computations.
To compute $A_{LD}$, we saturate $T(q)$ with the
$D,D^*$ intermediate states. These pole terms are then obtained in
terms of the $B \rightarrow D,D^{*}$
and $D \rightarrow K\pi $ and $D^{*} \rightarrow K\pi $ semi-leptonic decay form
factors at each vertex of the  pole diagrams and are given
by heavy quark effective theory and chiral effective lagrangian with the
form factors extrapolated to hard meson momenta. The strong
$D D^{*}\pi$ coupling constants with hard pion as in $B \rightarrow K\pi$
decay can be obtained by a suppression factor 
relative to the soft pion limit by
$G_{D^*D\pi}=\frac{2m_D \,g}{f_\pi}\,F(\vert \vec p_\pi\vert)$,
$F(\vert \vec p_\pi\vert)$ is 
normalized by $F(0)=1$ and $g\approx 0.4$ in the soft pion limit. For
$\vert\vec  p_\pi\vert \simeq m_B/2$,  
$ F(\vert \vec p_\pi\vert)~=~0.065\pm 0.035~$ in the constituent
 quark model.

Since the threshold for the $D_{s},D$ and   
$D_{s}^{*},D^{*}$ production  is below the $B$ meson mass, the $D_{s}$
and $D_{s}^{*}$ pole term for the $D,D^{*} \rightarrow K\pi$ form factors
have an absorptive part. This pole term  is in fact a rescattering  term
via the Cabibbo-allowed $B \rightarrow D_{s}D$ decays followed by the
strong annihilation process
$D_{s}D\rightarrow K\pi$ and can be obtained from  the unitarity 
of the $B\rightarrow K\pi$ decay amplitude. We have, for the $D_{s}D$ 
channel,
\be
 {\rm Disc}\, A_{LD}  =  -\frac{m_D}{16\pi^2
m_B}\, {\sqrt{\omega^{*2}-1}} \times \int d\vec n \, A(B\to
D_s D)\, A(D_s D\rightarrow K\pi)\ ,
\ee
and similar expressions for the $D_{s}^{*}D^{*}$ contribution. To
compute the absorptive part, we use factorization model which reproduces
well the $B\rightarrow D_s D $, $B\rightarrow D_s^{*} D^{*} $ decay rates.
$A(D_s D\rightarrow K\pi)$, $A(D_s^{*} D^{*}\rightarrow K\pi)$ are given by
the $t$-channel  $D,D^{*}$ exchange Born terms which are
$O(G_{D^*D\pi}^{2}) $. However the rescattering amplitudes 
$A(D_s D\rightarrow K\pi)$ and $A(D_s^{*} D^{*}\rightarrow K\pi)$ in the $B$ mass region,
being exclusive processes at high energy, should be
suppressed. This is taken into account by  the suppression factor
$F(\vert \vec p_\pi\vert) $ . As said earlier, the 
$A(B \rightarrow D_{s} D) $ and $A(B \rightarrow D_{s}^{*} D^{*}) $ amplitudes 
are  bigger than   $A(B \rightarrow K\pi)$  by a factor
$a_{2}/a_{6} \approx 20$ in factorization model. Hence
the absorptive part due to charmed meson rescattering effects
could be large. 
To find the real part, we compute  all Feynman diagrams for $T(q)$
and integrate over the virtual current momentum $q$ up to a cut-off
$\mu =m_{b}$.
It is possible to choose a cut-off momentum by a change of 
variable of variable $q=p_B - p_{D^{(*)}}$ to the  momentum $\ell$ 
defined by the formula
\begin{table}[hbt]
\caption{{\small  Theoretical values for $A_{\rm T+P}$ (Tree+Penguin
amplitude) and $A_{\rm ChP}$ (Charming Penguin amplitude).}}
\begin{center}
\begin{tabular}{|ccc|}
\hline
\hline {\rm Process} & $A_{\rm T+P}\times 10^8$ GeV  &  
$A_{\rm ChP}\times 10^8$ GeV \\
\hline
$  {B}^{+}\, \rightarrow \,K^{0}\pi^{+}$ & $+1.69$ & $+2.06\ +\ 2.36\,i$\\
 ${B}^{+}\, \rightarrow \,K^{+}\pi^{0}$ &
$+1.21\ -\ 0.498\,i$  & $ +1.45\ +\ 1.67\,i $ \\
 ${B}^{0}\, \rightarrow \, K^{+}\pi^{-}$ &
$+1.32\ -\ 0.634\,i$  & $+2.06\ +\ 2.36\,i$ \\
${B}^{0}\, \rightarrow \,K^{0}\pi^{0}$ &
$-0.921\ -\ 0.0497\,i$  & $-1.45\ -\ 1.67\,i $ \\
$ {B}^{+}\, \rightarrow \, \pi^{+}\pi^{0}$&
$-1.35\ -\ 1.79\,i $  & $ 0 $\\
${B}^0\, \rightarrow\, \pi^{+}\pi^{-}$ &
$-1.85\ -\ 2.16 \,i$  & $ -0.576\ -\ 0.648\,i $\\
${B}^{0}\, \rightarrow \,\pi^{0}\pi^{0}$&
$+0.0516\ +\ 0.379 \,i$& $ -0.576\ -\ 0.648\,i$\\
$ {B}^{+} \,\rightarrow\, K^{+} \eta$ &
$-0.0491\ -\ 0.415 \,i$  & $ +0.0830\ + \ 0.0896\,i  $\\
$ {B}^{+} \,\rightarrow\, K^{+}\eta^\prime$ &
$+1.40\ -\ 0.261 \,i$ & $ +2.53\ +\ 2.83\,i  $\\
${B}^{0} \,\rightarrow\, K^{0}\eta$ &
$~ +0.172\ -\ 0.0418 \,i$  & $ +0.0830\ +\ 0.0896\,i$\\
${B}^{0} \,\rightarrow\, K^{0}\eta^\prime$ &
$+1.54 \ -\ 0.0269\,i $ & $+2.53\ +\ 2.83\,i $\\
\hline
\hline
\end{tabular}
\end{center}
\end{table}

\begin{table}[hbt]
\caption{{\small  Theoretical values for the  CP averaged
Branching Ratios (BR). The last two columns are the more recent BR from
Babar\cite{Babar} and  Belle Collaboration \cite{Belle}. }}
\begin{center}
\begin{tabular}{|ccccc|}
\hline
\hline 
\kern -0.4cm {\rm Process} &\kern -0.3cm {\rm BR$\times 10^{6}\,$
\kern -0.2cm (\rm T+P)}&\kern -0.2cm {\rm BR$\times 10^{6}$\kern -0.2cm\, 
(\rm T+P+ChP)} &\kern -0.3cm {\rm BR$\times 10^{6}$(Babar)}\kern
-0.2cm & {\rm BR$\times 10^{6}$(Belle)}\kern -0,2cm \\
\hline
$B^{\pm}\rightarrow K^{0}\pi^{\pm}$ & $\sim 2.7 $ & $ 18.4 \pm\ 10.8 $
&  $17.5_{-1.7}^{+1.8}  \pm 1.3$ & $19.4_{-3.0}^{+3.1}  \pm 1.6$  \\
$B^{\pm}\rightarrow K^{\pm}\pi^{0}$ & $\sim 1.6 $ & $ 9.5 \pm\ 5.5  $
& $12.8_{-1.1}^{+1.2} \pm 1.0$  & $13.0_{-2.4}^{+2.5}  \pm 1.3$  \\
$B^{0} \rightarrow K^{\pm}\pi^{\mp}$ & $\sim 1.9 $ & $ 15.3 \pm\ 9.9 $ & 
$17.9 \pm 0.9\pm 0.7$ & $22.5_{-1.8}^{+1.9}  \pm 1.6$\\
$B^0 \rightarrow K^{0}\pi^{0}$ & $\sim 0.75 $ & $ 7.4 \pm 4.8 $ & 
$10.4 \pm 1.5 \pm 0.8$ &  $ 8.0_{-3.1}^{+3.3}  \pm 1.6$  \\
$B^{\pm}\rightarrow \pi^{\pm}\pi^{0}$ &$\sim 4.8 $ & $  \sim 4.8 $ & 
$5.5_{-0.9}^{+1.0}  \pm 1.3$ & $7.4_{-2.2}^{+2.3}  \pm 0.9$ \\
$B^0 \rightarrow\pi^{+}\pi^{-}$ & $\sim 7.2 $ & $ 9.7 \pm 2.3 $ & 
$4.6\pm 0.6 \pm 0.2$ & $5.4\pm 0.12 \pm 0.5$ \\
$B^0 \rightarrow \pi^{0}\pi^{0}$ & $\sim 0.06 $ & $ 0.37\pm 0.35 $ 
&$ < 3.6$  & $ < 6.4$ \\
\hline\hline
\end{tabular}
\end{center}
\end{table}
\be
q=p_B - p_{D^{(*)}}\equiv (m_B-m_{D^{(*)}}) v -\ell\ .
\ee
As discussed in \cite{Isola2}, the chiral symmetry breaking scale is
about $1\rm\,GeV$ and the mean charm quark momentum $k$ for the on-shell
$D$ meson is about $300\rm \, MeV$, the virtual momentum $\ell$ should be
below $0.6\rm \, GeV$, hence 
a cut-off $\mu_\ell\approx 0.6~{\rm GeV}~$. The real part is then given
by a Cottingham formula as follows (the notations are those 
of \cite{Isola2}) .
\bea
 {\rm Re}A_{LD}&=&\frac{i\,K}{2\ (2\pi)^3}
\int_0^{\mu_\ell^2} dL^2                              
\int_{-\sqrt{L^2}}^{+\sqrt{L^2}} dl_0
\sqrt{L^2-l_0^2}    \cr
&&\times \int_{-1}^{1}d \cos(\theta)\ i      
 \, \left\{\frac{j_D^\mu\ h_{D\, \mu} }{p_D^2 - m_D^2} +
\frac{\sum_{pol}\ j_{D^*}^\mu\ h_{{D^*}\, \mu} }{p_{D^*}^2 -
m_{D^*}^2} \right\}.
\eea
The results of the calculations are shown in Table 1
taken from \cite{Isola}. The short-distance part $A_{\rm T+P}$ is obtained
with $c_2=1.105,~c_1=-0.228,~c_3=0.013,~c_4=-0.029,~c_5=0.009,
~c_6=-0.033 $ \cite{Buras1}, $\gamma =54.8^{\circ}$. 
$F_0^{B\, M'}(m^2_M)\approx F_0^{B\, M'}(0)=0.25$ 
($M,~M'=K,~\pi^\pm$) given by QCD sum rules \cite{Colangelo}, and 
is smaller by $20-30\%$ than the lattice and BWS values \cite{Ali}. 
As seen from Table 1, the charming penguin term $A_{\rm ChP}$
and its absorptive part are comparable to the short-distance
part. The predicted 
branching ratios for $B \rightarrow K\pi$ and $B \rightarrow \pi\pi$ decays are shown in
Table 2 with results from Babar\cite{Babar} and 
Belle\cite{Belle}. We note general agreement with data
considering various uncertainties on the value for the cut-off $\mu_{\ell}$ 
and the suppression factor 
$F(\vert \vec p_\pi\vert) $  as discussed in \cite{Isola2,Isola}. 
The $K\eta^{\prime}$ modes are also enhanced  by the charming penguin 
contribution as shown in Table 1 \cite{Isola}. 

The CP asymmetry is
${\cal A}_{\rm CP} = (\bar{\Gamma} - \Gamma|)/(\bar{\Gamma} + \Gamma)$.
Since $\bar{\Gamma} - \Gamma$ is given by
$4\,\rm Im \,A_{\rm T+P}\,Im \,A_{\rm ChP} $ which
depends only on $\rm Im \,A_{\rm T +P}$.
and the absorptive part of the charming penguin, the absolute asymmetry
$|{\cal A}_{\rm CP} |$ suffers from less uncertainties than
the real part in our calculations. From the Table 1, we obtain 
for $|{\cal A}_{\rm CP} |$, the value 0.16, 0.17 and 0.25 for the 
mode $K^{+}\pi^{0}$, $K^{+}\pi^{-}$ and $\pi^{+}\pi^{-}$, respectively,
This is in contrast with  small CP asymmetries predicted by QCD-improved 
factorization \cite{Beneke,Du,Muta} for these decays. 
Our values are similar to those given in \cite{Ciuchini2,Ciuchini3}
and compatible with the measured asymmetries by  Babar\cite{Babar}
and Belle \cite{Belle} Collaborations who found smaller asymmetries
but with large error. One exception is the Babar value 
${\cal A}_{\rm CP} =-0.102\pm 0.05\pm 0.016$ for the mode $K^{+}\pi^{-}$ 
which is consistent with the absolute value $|{\cal A}_{\rm CP}|=0.17$ 
we found. 

In conclusion, long-distance and charmed meson 
rescatterings effects are capable of producing large $B \rightarrow K\pi$
decay rates and CP asymmetries. More data from Babar and Belle
will tell us if these inelastic FSI rescattering effects are 
indeed present in charmless $B$ decays, a rare situation 
in nonleptonic decays.  

\bigskip

I would like to thank S. Narison and the organisers of QCD02 for the
warm hospitality extended to me at Montpellier.

\end{document}